# Unveiling delay-time-resolved phase noise dynamics of narrow-linewidth laser via coherent optical time domain reflectometry

Liang Zhang[*], Liang Chen, Xiaoyi Bao[§]

*Department of Physics, University of Ottawa, Ottawa, Ontario K1N 6N5, Canada*

## ABSTRACT

**Laser with high spectral purity plays a crucial role in high-precision optical metrology and coherent communication. Thanks to the rapid development of laser frequency stabilization, the laser phase noise can be remarkably compensated, allowing its ultra-narrow linewidth subject to mostly quantum limit. Nevertheless, the accurate characterization of phase noise dynamics and its intrinsic linewidth of a highly coherent laser remains ambiguous and challenging. Here, we present an approach capable of revealing delay-time-resolved phase noise dynamics of a coherent laser based on coherent optical time domain reflectometry (COTDR), in which distributed Rayleigh scattering along a delay fibre essentially allows a time-of-flight mapping of a heterodyne beating signal associated with delay-time-dependent phase information from a single laser source. Ultimately, this novel technique facilitates a precise measurement of ultra-narrow laser linewidth by exploiting its delay-time-resolved phase jitter statistics, confirmed with the analytical modelling and numerical simulations.**

Correspondence should be addressed to L. Z. ([*]email: opticaliang@gmail.com ) or to X. B. ([§]email: xbao@uottawa.ca )

Coherent laser sources with highly purity spectrum are at the heart of high-precision measuring science, including laser interferometer gravitational-wave observatory (LIGO)[1], optical atomic clock[2] and high-resolution spectroscopy[3]. Versatile applications such as coherent commumication[4,5], narrow-linewidth microwave/terahertz photonic generation[6,7] also demand an ultra-narrow linewidth laser involving squeezed phase noise to upgrade their performance towards a fundamental limit. A rapid development of frequency stabilization technique currently enables a coherent laser radiation with an extremely narrow linewidth down to mHz[8-13], however, the characterization of such narrow linewidth remains challenging.

The linewidth of a laser is fundamentally governed by the laser radiation coupled with a quantum process of spontaneous emission (i.e., pure white noise), namely "Shawlow-Townes linewidth"[14], which theoretically exhibit a lorentzian lineshape[15]. As a crucial parameter, this intrinsic quantum-limit linewidth generally reflects the short-term stability as well as the temporal coherence of a laser. It can be evaluated by the heterodyne detection of beating notes superimposed with another identical laser, which is however not usually available in a breeze, particularly for an elaborately stabilized laser system. Alternatively, a delayed self-heterodyne interferometer (DSHI) approach, in which one portion of de-correlated laser beam after a long delay time is obligatorily required to replace the 2nd identical laser[16], is proposed and intensively employed to characterize the linewidth as well as the phase noise [17-20]. However, in spite of its simplicity, the DSHI approach turns out to be physically incompetent for an ultra-narrow linewidth (e.g. < 1 kHz) due to a remarkable attenuation (>40 dB) over hundreds of kilometres delay fibre even though at a minimum loss spectral window around 1.55 μm. This challenge was overcome in part by a loss-compensated recirculating technique with an extended long delay[21-23] or the strong coherent envelope analysis associated to a relative short delay[24-26]. However, its robustness remains elusive because of a definite delay-time-dependent behaviour in all above-mentioned approaches[27]. More importantly, the inevitable 1/f frequency noise over a long delay beyond the laser coherence time imposes a deviation of the intrinsic linewidth from a lorentzian shape to a Gaussian broadening one[28-30], leading to a dilemma for an intrinsic linewidth characterization. On the other hand, interferometric recovery of phase or frequency noise of the laser by means of a short-delay-time interferometer are theoretically[31] and experimentally[32-36] demonstrated to estimate the intrinsic laser linewidth albeit with a non-straightforward manner, however, these rather sophisticated measurement systems place additional stringent constrictions, resulting in either an over-optimistic estimation of the intrinsic linewidth[37] or hindering the flexibility for practical implementation. Consequently, a precise characterization of the fundamental laser linewidth subject to a quantum-limit phase noise dynamics is highly demanded not only being oriented by practical applications but also in a more fundamental prospective of laser physics.

In this communication, we demonstrate a novel technique to reveal the phase jitter dynamics of a coherent laser by employing a coherent optical time domain reflectometry (COTDR), for the first time to the best of our knowledge, in which distributed Rayleigh scattering along a segment of delay fibre technically introducing a time-of-flight for the heterodyne beating note carrying a delay-time-resolved phase information of the laser under test. A fading-free phase jitter demodulation based on Pearson correlation coefficient (PCC) is developed, which allows a statistical analysis of phase noise dynamics, agreeing well with theoretical analysis and numerical simulation. In this scenario, the intrinsic linewidth is ultimately characterized by means of the statistics of the delay-time resolved phase jitter. The proposed approach breaks through the dilemma of narrow linewidth characterization in a conventional DSHI approach, highlighting promising potentials in fundamental laser physics and coherent communication.

## RESULTS

### Laser intrinsic linewidth modelling

Spontaneous emission as a quantum process is fundamentally ubiquitous during a laser generation, which primarily introduces the phase fluctuation of the laser field in a random fashion and deviates its output electrical field from an ideal sinusoidal wave. A quasi-monochromatic laser field can be modelled as a sinusoidal wave with random fluctuations of the phase:

$$E(t) = E_0 e^{j[2\pi f_0 t + \varphi(t)]} .$$ (1)

where $E_0$ is a stationary value for the amplitude, $f_0$ is the nominal central optical frequency, respectively. The white frequency noise-induced phase fluctuation $\varphi(t)$ is a continuous random walk, which can be modelled by the Wiener process[38]. Thus, phase jitter after a *positive* time delay $\tau$ can be defined as $\Delta_\tau \varphi(t) = \varphi(t + \tau) - \varphi(t)$, which is a random variable with a probability density function of a Gaussian distribution,

$$p(\Delta_\tau \varphi) = \frac{1}{\sqrt{2\pi\sigma_\varphi^2}} e^{-\frac{\Delta_\tau \varphi^2}{2\sigma_\varphi^2}} .$$ (2)

$\Delta_\tau \varphi$ has zero mean and its variance $\sigma_\varphi^2$ is linearly proportional to the time delay $\tau$. Theoretical results reveal that the relation between the intrinsic linewidth and the phase jitter variance [39,40],

$$\Delta \nu_c = \frac{\sigma_\varphi^2}{2\pi\tau} .$$ (3)

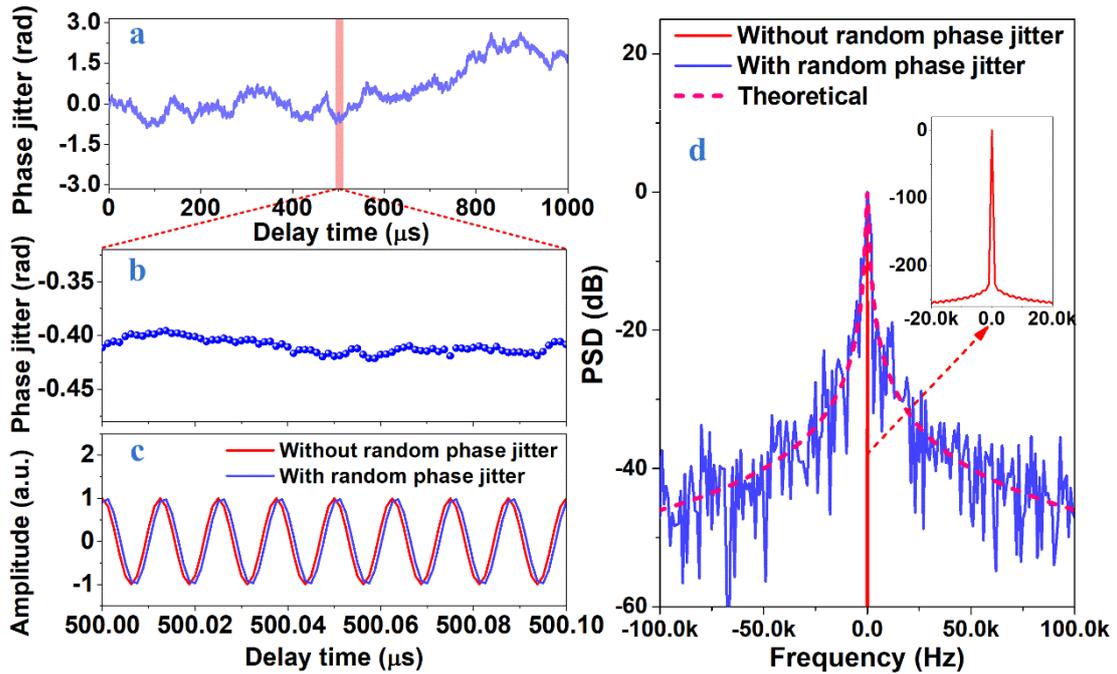

**Figure 1: Simulation of laser phase noise and intrinsic linewidth.** (a) Simulated random-walk phase noise by Wiener process, in which the phase jitter variance is given by $\sigma_\varphi^2 = 2\pi\Delta\nu_c\tau$ with $\Delta\nu_c$ of 1 kHz. (b) and (c) are zooming in results of random phase noise and corresponding electrical field of beating signals with (blue) and without (red) random phase jitter. Here, the beating signal originates from a self-heterodyne of the laser field with a frequency shift of 80MHz; (d) Power spectral density (PSD) of the beating signals with (blue) and without (red) random phase jitter. The peak is normalized to a maximum of 0 dB, and the center frequency is normalized to 80 MHz, which is in agreement with theoretical lorentzian laser line shape (magenta dash line) with a full-width-half-maximum of 1 kHz subject to quantum-limit phase noise.

Given a laser with a lorentzian linewidth of 1 kHz, a random walk of the laser phase jitter can be simulated by Wiener process with computational random numbers, as shown in Fig. 1(a). The electrical field of the laser can be recalculated by adding the random phase jitter into its original pure sinusoidal waves. Then, the laser linewidth can be determined by the 3-dB bandwidth of the power spectral density of the laser spectrum. As shown in Fig. 1(d), the linewidth of the laser with quantum noise-induced random phase jitter was broadened to a Lorentz line shape with a 3-dB linewidth of 1-kHz.

## Laser phase noise in COTDR system

Coherent optical time domain reflectometry (COTDR) is an intriguing technique for characterization and fault location in optical fibre transmission systems[41,42] as well as distributed fibre-optic dynamic strain sensing aiming to structural health monitoring [43,44]. In coherent OTDR system, the lightwave from a narrow-linewidth laser source is divided into two paths: the lightwave in the first path is temporally modulated into a pulse with a central frequency shift $f_s$ and then launched into one-end of long fibre while the lightwave in the second path remains *cw* as the local reference. As the light pulse injects into the optical fibre with numerous randomly localised Rayleigh scatters originating from inhomogeneities of the refractive index, a backscattered coherent speckle signal can be formed as the vector sum of these back reflections. Then, the Rayleigh backscattered light beams are recaptured and superimposed with the local reference. By ignoring the polarization mismatch, the electrical field of the beating note can be expressed as,

$$E_b(t) = E_1 e^{j[2\pi f_o t + \varphi(t)]} + E_2 e^{j[2\pi(f_o+f_s)t + \varphi(t+\tau_z) + \phi_R(z)]} \qquad (4)$$

where the $\varphi(t)$ is the laser phase noise and $\phi_R(z)$ is the Rayleigh-induced additional phase shift due to the incident light at the scatter location $z$ along the fibre. Assuming the Rayleigh fibre with an approximate constant group refractive index along the fibre direction $n_g(x) \approx n_0$ (i.e. $n_0 \sim 1.5$ in standard single mode fibre), the delay time $\tau_z$ corresponding to a location $z$ can be given by

$$\tau_z = \frac{2}{c}\int_0^z n_g(x)dx \ \approx \frac{2n_0}{c}z. \tag{5}$$

By removing the DC component and down-converting the frequency into the baseband of a balanced photodetector (BPD), the detected heterodyne beat-note can be described as,

$$I_{ac}(t) \propto \ 2E_1E_2\cos\bigl(2\pi f_s t + \Delta_\tau\varphi(t) + \phi_R(z)\bigr). \tag{6}$$

where $\Delta_\tau\varphi(t) = \varphi(t + \tau_z) - \varphi(t)$ is the phase jitter of the laser source under the time delay $\tau_z$. Note that, the random nature of Rayleigh backscattering leads to non-uniform phase shift $\phi_R(z)$ along the fibre, however, $\phi_R(z)$ can be approximately constant at the "frozen" Rayleigh scatters location $z$ along the fibre by the removal of any external thermal or acoustic disturbance. Consequently, the COTDR trace pattern $I(t)$ is naturally accompanied by a delay-time-dependent phase information of the laser source. According to Eq. (5), the delay time $\tau$ can monotonously vary from 0 to $z$ along the distributed Rayleigh fibre direction.

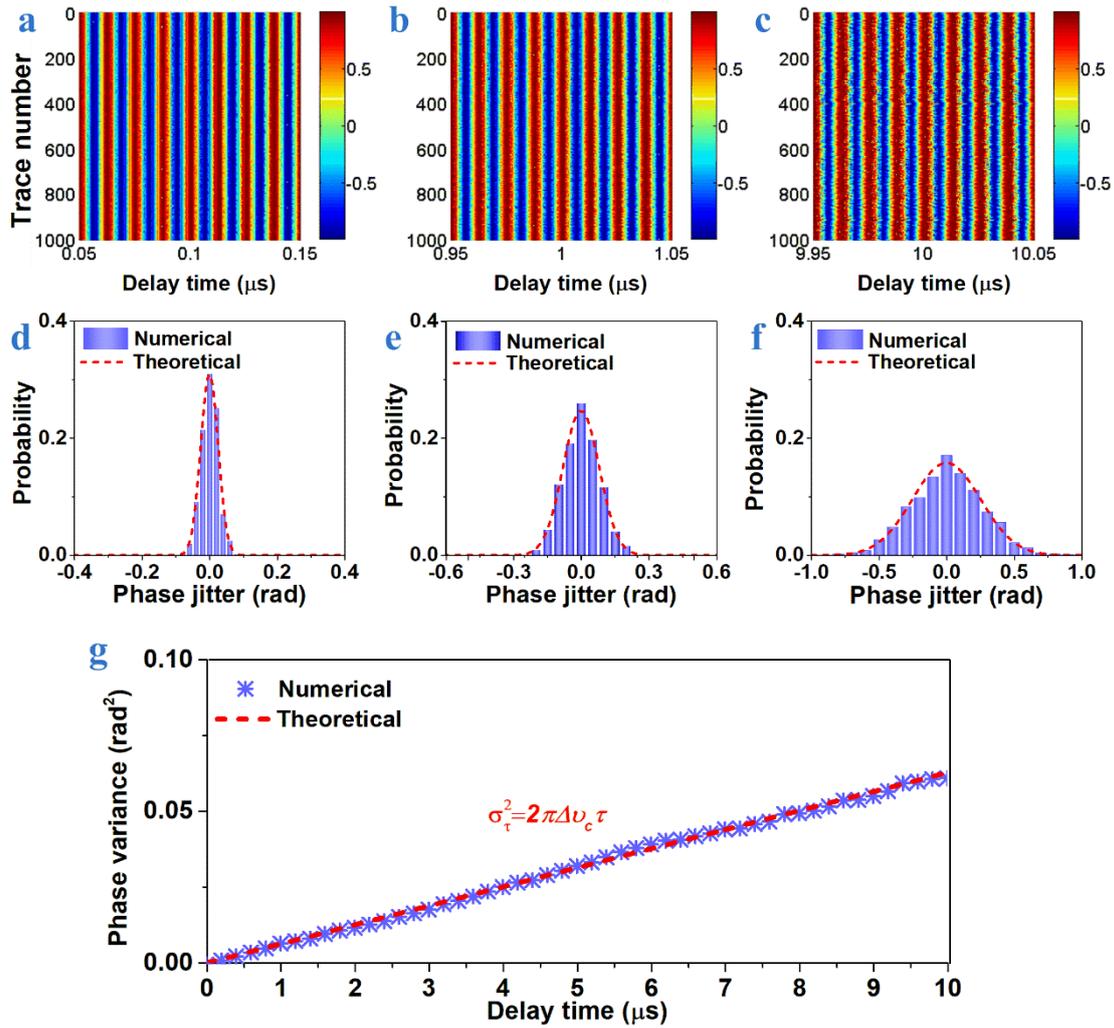

**Figure 2: Simulation of the phase noise dynamics and statistics of a coherent laser with a quantum-limit linewidth of 1-kHz.** (a) ~ (c) show numerically simulated evolutionary trajectories of 1000 heterodyne beating notes in COTDR at delay time of 0.1 μs, 1 μs and 10 μs. Random phase noise of laser with 1-kHz intrinsic linewidth was numerically generated by a Wiener process from random number generation. Note that, Rayleigh scattering-induced $\phi_R(z)$ in each fiber location $z$ was setting as null for simplicity without loss of generality. (d) ~ (f) represent corresponding probability histograms (blue bar) of the phase jitter at delay time of 0.1 μs, 1 μs and 10 μs, respectively; (g) Numerical phase jitter variance as a function of delay time (blue star dots), corresponding well with the theoretical curve (red dash line) given by $\sigma_\varphi^2 = 2\pi\Delta\nu_c\tau$.

According to Eq. (3), the intrinsic linewidth is theoretically determined by the delay-time-dependent phase jitter variance. In Fig. 2, we simulate the COTDR-based heterodyne beating notes of a laser with 1-kHz intrinsic linewidth at different delay times of 0.1, 1, 10 μs, following random phase noises by repeating 1000 times Wiener process computation. It can be seen that, the pattern of beating notes at 0.1-μs delay time exhibits a minor variation over 1000 repeating cycles while larger ripples are significantly introduced as the delay time increases to 10 μs, indicating a worse phase jitter variance. Correspondingly, the delay-time-dependent phase jitter variance $\sigma^2$ can be statistically obtained at each delay time. In Fig. 2 (g), numerical results show that the phase jitter variance is proportional to the delay time with a linear slope of $2\pi\Delta\upsilon_c$, which is in agreement with the theoretical analysis.

## Fading-free retrieval of delay-time-resolved phase jitter

In typical COTDR, the random nature of Rayleigh backscattering as well as the random state of polarization of the backscattered signals along the fibre occasionally deteriorates the heterodyne beating notes, leading to a notorious signal fading[45] which is detrimental to the phase recovery. To address it, we develop a fading-free approach based on Pearson Correlation coefficient (PCC) for the phase jitter retrieval in COTDR. Figure 3 shows the schematic diagram to demodulate the delay-time resolved phase jitter variance by statistical analysis with assistant of COTDR. By launching two pulses with a time interval of $\Delta t = t_j - t_k$, a pair of coherent COTDR traces ($I_j$, $I_k$) carrying the delay-time-dependent phase from the laser under test can be acquired. Taking into account of cosine function type of beating note traces, the absolute value of the phase jitter at delay time $\tau$ can be deduced by (See **Supplement S1**),

$$|\Delta_\tau\varphi| = \cos^{-1}\big(\rho(\tau)\big) \tag{7}$$

Here, $\rho$ is the calculated PCC between two beating note traces at the central delay time of $\tau_z$ within a small time window width of $t_w$,

$$\rho|_{\tau=\tau_z} = \mathbb{PCC}\left\{I_j\left(\tau_z - \frac{t_w}{2}, \tau_z + \frac{t_w}{2}\right), I_k\left(\tau_z - \frac{t_w}{2}, \tau_z + \frac{t_w}{2}\right)\right\}. \tag{8}$$

It is important to mention that, the PCC-based phase jitter retrieval with a moderately small time window (~100 ns) basically guarantees a negligible phase variation within this small time window, particularly for a highly coherent laser source. For instance, for a 1-kHz laser, the corresponding phase jitter variance within a timescale of 100 ns is around $2\pi \times 10^{-4}$ rad$^2$. In this scenario, the fading problems in COTDR, caused by the destructive interference of random Rayleigh scattering as well as the polarization mismatch between the backscattered signal and the local reference[45], can be alleviated to a large extent since the occasional ultra-weak signal fading occurs at certain points within single pulse width cell[44]. Eventually, the proposed PCC-based approach turns out to be effective for the fading-free phase jitter retrieval.

By launching a number of optical pulses and repeating *Ns* times beating note traces acquisition, the variance of the absolute phase jitter $|\Delta_\tau\varphi|$ at the delay time of $\tau_z$ could be statistically calculated as,

$$\sigma'^2_\tau|_{\tau=\tau_z} = \sum_{j=1}^{j=Ns}\left(|\Delta_\tau\varphi_j| - \langle|\Delta_\tau\varphi_j|\rangle\right)^2|_{\tau=\tau_z}. \tag{9}$$

It is worth mentioning that, the time interval *Δt* between two calculated beating note traces should be longer than the coherent time of the laser in order to mitigate any correlation between them for independent statistical analysis of the phase jitter variance. In this way, the delay-time resolved phase jitter variance can be mathematically obtained (See **Supplement** S2 for the derivation),

$$\sigma^2_\tau = \frac{\sigma'^2_\tau}{2\left(1-\frac{2}{\pi}\right)}. \tag{10}$$

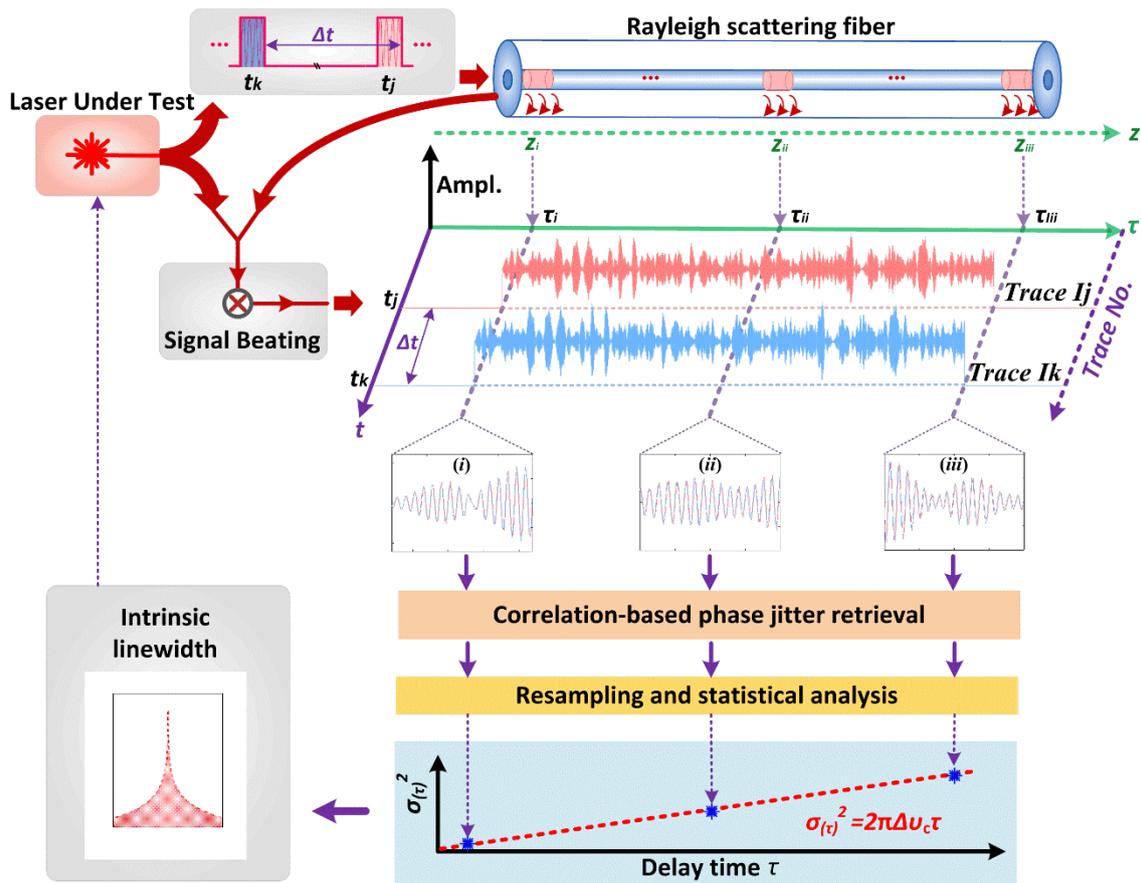

**Figure 3 Schematic diagram of delay-time-resolved phase jitter retrieval and statistical analysis relying on COTDR. In COTDR system, optical pulses modulated from the laser under test are consecutively reflected by distributed Rayleigh scattering along optical fibre, allowing a time-of flight mapping of the heterodyne beating signal associated to delay-time-dependent laser phase noise. By launching a number of optical pulse pairs with a time interval larger than the laser coherent time ($\Delta t > t_c$), the phase jitter at different delay time $\tau$ can be extracted by employing the Pearson Correlation Coefficient (PCC) method in a fading-free fashion. The extracted phase jitter can be then statistically processed to reveal a delay-time-resolved phase noise dynamics of the laser, which further reflects the intrinsic linewidth of the laser under test.**

## Experimental setup

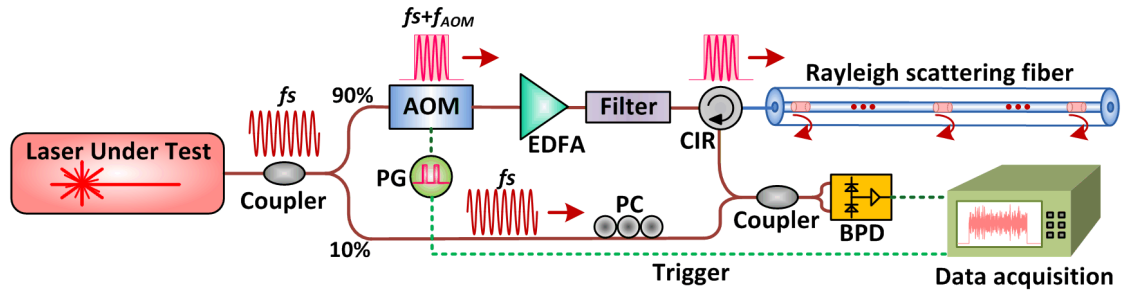

**Figure 4 Experimental setup for the delay-time resolved measurement of laser phase jitter. (Component label abbreviations: AOM—Acoustic-optic modulator, PG—Pulse generator, EDFA—Erbium doped fibre amplifier, CIR—optical circulator, PC—polarization controller, BPD—balanced photodetector.) The delay fibre is 1-km standard single mode fibre (SMF) which was placed in a sound-proof box for ambient acoustic isolation. The symbols of $f_s$ and $f_{AOM}$ represent the center frequency of the laser source and the frequency shift induced by the AOM, respectively.**

As depicted in Fig. 4, the lightwave beam from the laser under test is split into two branches through an optical coupler. The upper branch was modulated into an optical pulse through an acoustic-optic modulator (AOM) with a frequency-shift ($f_{AOM}$), which is driven by a pulse generator (PG) with a pulse width of 100 ns and a pulse repetition rate of 10 kHz. The generated optical pulses are amplified by an EDFA and then pass through a narrowband filter to suppress undesired noises from amplified spontaneous emission. By launching optical pulses into the delay fibre through an optical circulator (CIR), the lightwave is backscattered by numerous distributed Rayleigh scatters along the delay fibre by introducing the time-of-flight for variable delay times. The 10% lower branch is used as the local oscillator (LO) and to superimpose with backscattered signals through another 50/50 coupler. A polarization controller (PC) is inserted in order to optimize the polarization matching between the LO and scattered signals. Afterwards, the beating signals was converted into electrical signals by a balanced photo-detector (BPD) (PDB130C, Thorlabs) and then digitized by a high-speed oscilloscope (MSOS804A, Keysight). Here, the balanced heterodyne detection is commonly adopted to reduce the dynamic range requirement of the detector as well as improve the sensitivity [42].

## Delay-time-resolved laser phase noise dynamics and statistics

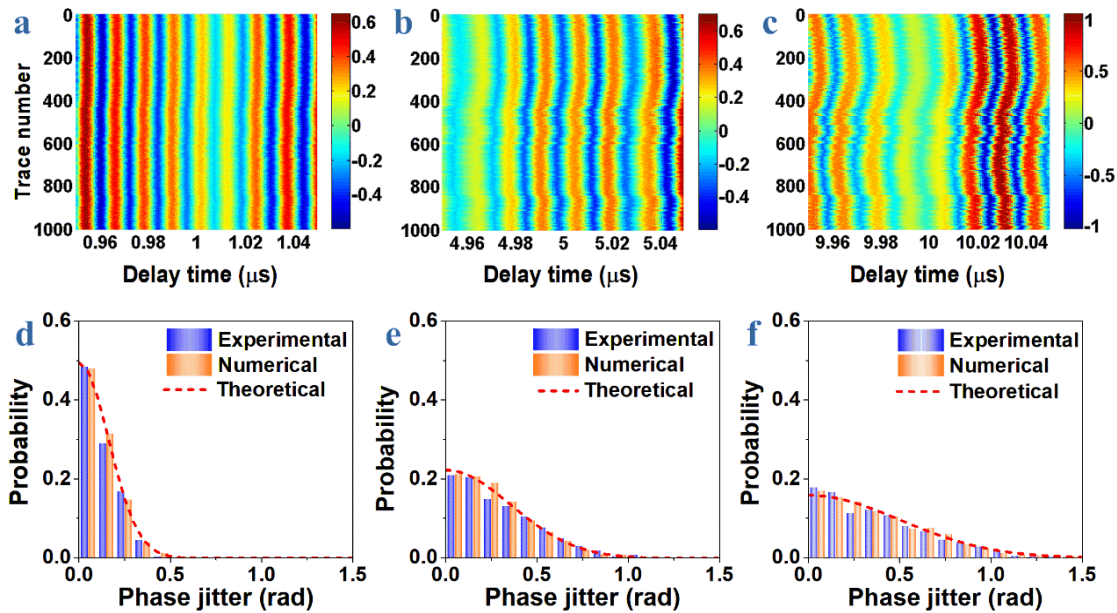

**Figure 5: Phase noise dynamic and statistics retrieval of a 2-kHz External Cavity Laser. (a)-(c) Evolutionary trajectory of 1000 beating notes at different delay times of: (i) 1 μs, (ii) 5 μs, (iii) 10 μs. (The time window $t_w$ is 100 ns for each delay time case). (d)-(f) show the experimental probability histograms (Blue) of the corresponding resolved absolute phase jitter $|\Delta_\tau \varphi|$, compared to numerical (Orange) and theoretical (Red dash line) simulation results.**

A 1550 nm External Cavity Laser (ECL) (ORION Module, RIO) with a narrow linewidth of 2 kHz was utilized as the laser under test. Figures 5(a)-(c) display the recorded beating note trace evolutions within a time window of 100 ns by consecutively launching 1000 optical pulses with a repetition rate of 10 kHz in COTDR system. It can be seen that the beating notes pattern show a small trace-by-trace variation at 1-μs delay time while the increased random phase noise apparently degenerates the beating note pattern as the delay time increasingly reaches 10 μs, which is coincident as the theoretical prospection. With the PCC-based phase jitter retrieval method, the absolute phase jitter $|\Delta_\tau\varphi|$ at delay time $\tau$ can be obtained through data processing. Considering the estimated coherent time of the 2-kHz laser (i.e., $t_c \sim \frac{1}{\Delta v_c} = 0.5\ ms$), two beating note traces with a time interval $\boldsymbol{\Delta t} = 0.6$ ms (e.g., the trace pairs of the 1st/7th , the 2nd/8th, …, the 994th /1000th) are consecutively selected to recover the phase jitters for statistical analysis. As illustrated in Figs. 5(d)-(f), the probability of the retrieved absolute value of the phase jitter $|\Delta_\tau\varphi|$ at delay times of 1, 5, 10 μs are statistically obtained, showing consistent matching with numerical and theoretical simulations.

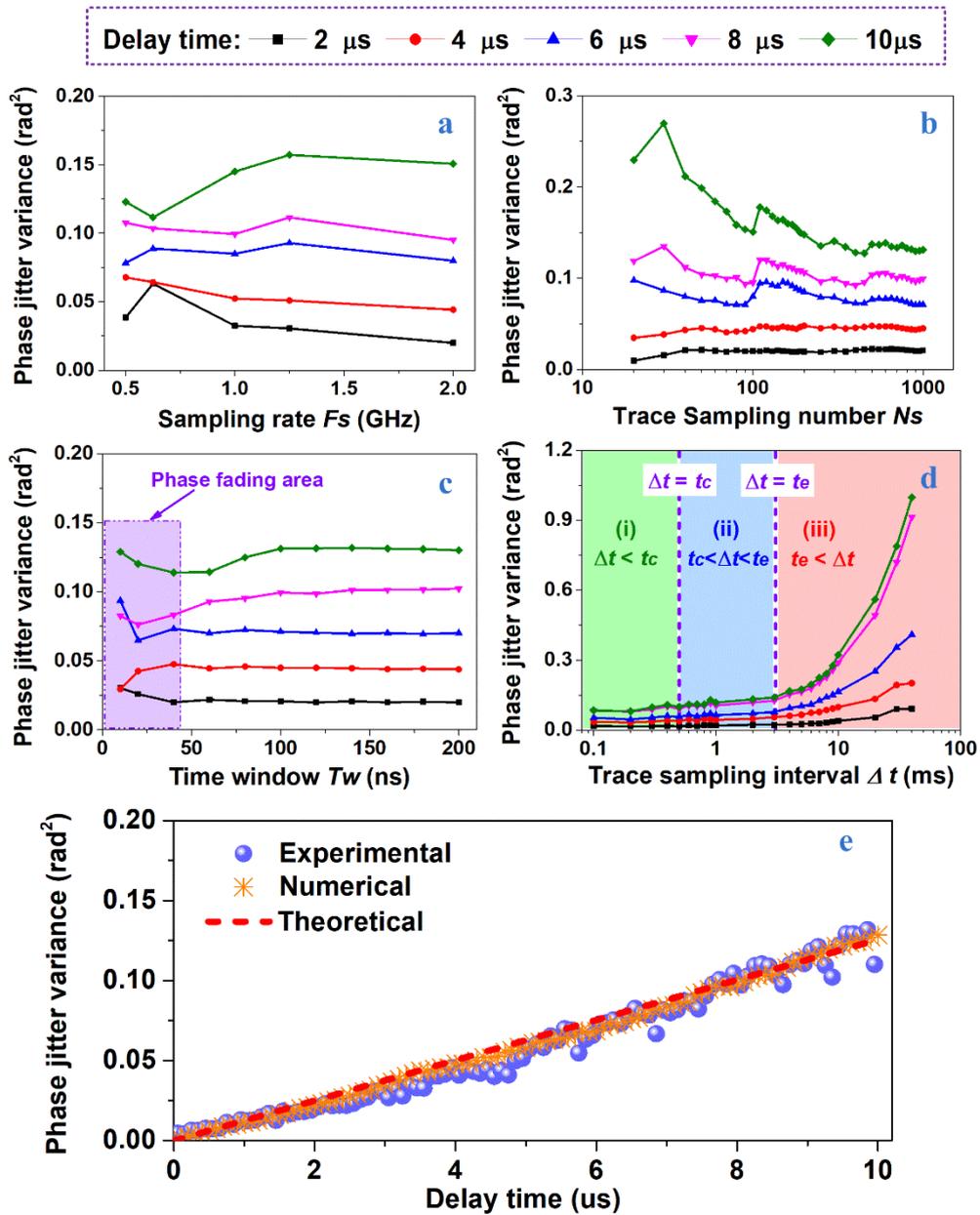

**Figure 6: Delay-time resolved phase jitter variance. Experimental investigation of data acquisition parameters: (a) Sampling rate *Fs*, (b) Trace sampling number *Ns*, (c) Time window *tw*, (d) Trace time interval *Δt*. Discrete delay times were selected as 2μs (black), 4μs (red), 6μs (blue), 8μs (megenta), 10μs (green). (e) Delay-time-resolved phase jitter variance. (Parameters setting as *Fs*=2GSa/s, *Ns* = 1000, *tw*=100ns and *Δt*=0.6ms.) Numerical (Orange star dots) and theoretical (Red dash line) results were implemented based on an identical laser linewidth of 2 kHz for comparison.**

The accuracy of statistically demodulated phase jitter variances would be affected by several factors in data acquisition and processing, such as the sampling rate ($Fs$) of the data acquisition, total beating note trace sample number ($Ns$), the time window ($t_w$) for absolute phase jitter recovery and the time interval ($\Delta t$) between two beating note trace samples. As shown in Fig. 6(a), the measured phase jitter variance at different delay times decrease as we increase the sampling rate of the data acquisition and remain stable at a sampling rate larger than 1 GSa/s. In statistical analysis, the calculated phase jitter variances were convergent as the sample number increasingly reached 1000, as depicted in Fig. 6(b). To go around of fading problems in COTDR system, a time window $t_w$ was properly chosen to demodulate the absolute phase jitter. In Fig. 6(c), the fading-induced error turns to be significant at a time window of less than 50 ns while the recovered phase jitter remained stable as $t_w$ exceeds 100 ns, implying an excellent performance in signal fading error suppression. In statistical analysis of the phase jitter variance, two consecutive trace samples should be independent which can only be guaranteed by an adequate time interval ($\Delta t > t_c$) between two optical pulses modulated from the same laser source. As can be seen in Fig. 6(d), smaller phase jitter variances can be found as the trace time interval is less than the coherent time of the 2-kHz laser ($t_c \sim {1}/{\Delta v_c} = 0.5\ ms$). The recovered phase jitter variances saturated at constant levels for different delay time cases at $\Delta t > 0.5$ ms. On the other hand, the phase jitter variance increased divergently as the trace time interval is longer than several milliseconds (~3 ms, defined as $t_e$) in our experiments, particularly at a larger delay time. It mainly attributes to additional 1/f frequency noise caused by the external disturbance imposed from the thermal and acoustic noise of surrounding environment [46], which primarily dominates in a low frequency domain of <100 Hz (corresponding to a time interval of

~10 ms). Ultimately, the delay-time resolved phase jitter variances of a narrow-linewidth laser were retrieved with the removal of any fading problems, in accordance with numerical and theoretical simulations, as shown in Fig. 6 (g). Note that, the phase jitter variance with finer delay time step can be quasi-continuously retained, albeit with the expense of computational budget.

## Intrinsic linewidth characterization of a highly coherent laser

The above-mentioned technique evidently reveals the phase noise dynamics of a coherent laser by statistically retrieving its delay-time resolved phase jitter variance. One of its potential applications can be explicitly utilized to characterize the quantum-noise-induced linewidth of a laser source, particularly for ultra-narrow linewidth of <1 kHz. According to Eq. (3), the white-noise-induced laser intrinsic linewidth can be readily deduced by the slope efficiency of the phase jitter variance with respect to the delay time in a straightforward manner.

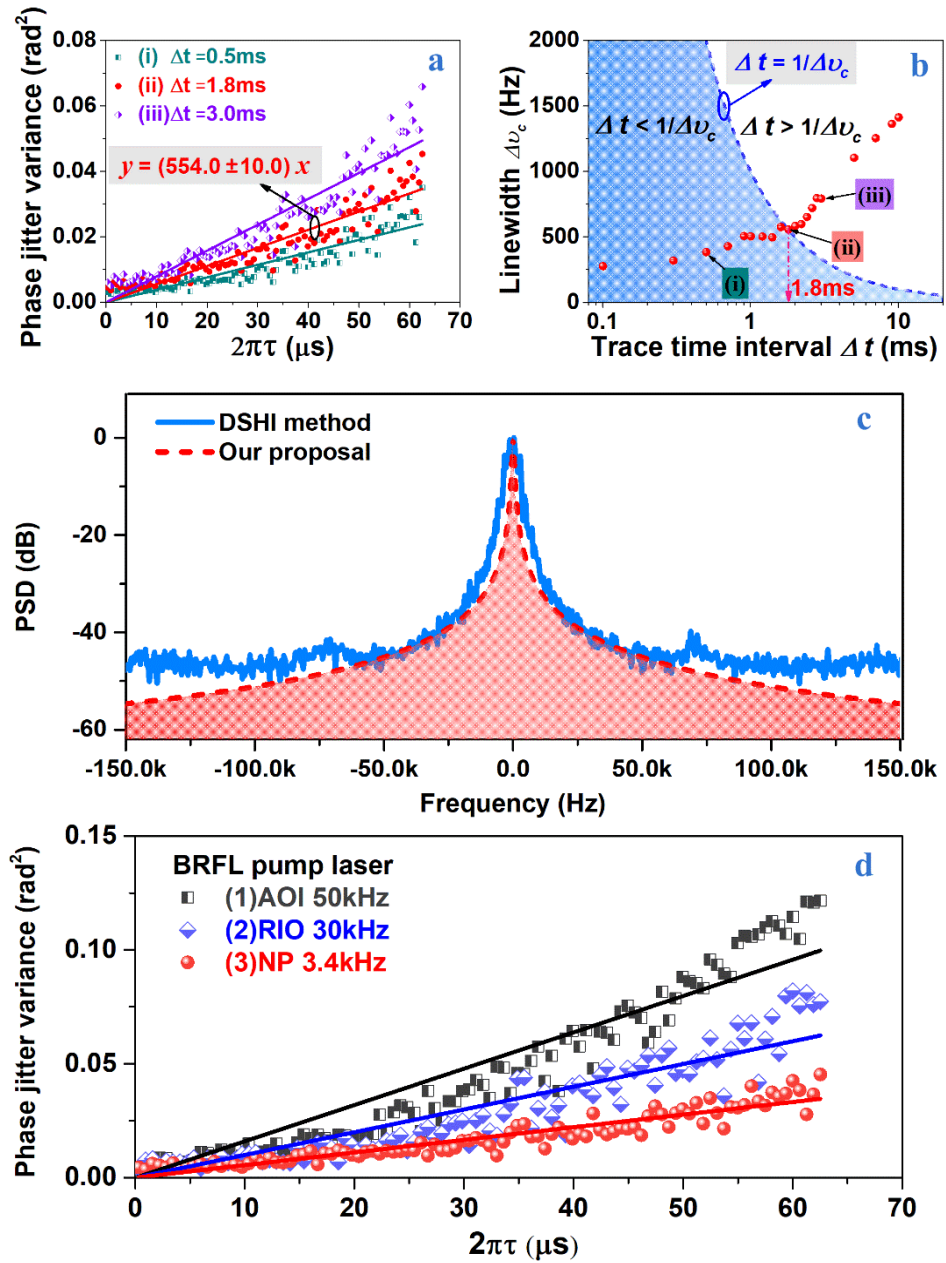

**Figure 7: Phase noise dynamics and intrinsic linewidth characterization of a sub-kHz Brillouin random fiber laser (BRFL).** (a) Phase jitter variance as a function of the delay time with the beating note trace interval of (i) 0.5 ms, (ii) 1.8 ms, (iii) 3.0 ms; the delay time is modified by multiplying a factor of $2\pi$ to facilitate the linewidth fitting. (b) Linewidth estimation as a function of trace time interval ranging from 0.1 ms to 10 ms. (c) Recovered laser spectrum PSD with a lorentzian linewidth estimated by our proposal in comparison to the DSHI results with 200-km delay fibre (See Methods). (d) Investigation of BRFL pump laser impact on phase noise dynamics and intrinsic linewidth. The BRFL was pumped by three different pump lasers: (1) AOI laser with 50 kHz (DFB, AOI Inc.), (2) RIO laser with 30 kHz (RIO009x, RIO Inc.) and (3) NP laser with 3.4 kHz (Rock Module,

**NP Photonics Inc.). Accordingly, the intrinsic linewidth of the BRFLs with three pump lasers were mathematically obtained as 554.0(±10.0) Hz, 998.1(±23.0) Hz, 1600.0(±30.2) Hz, respectively.**

For the proof-of-concept demonstration, a sub-kHz Brillouin random fibre laser (BRFL)[47] was investigated by our proposed approach. As a new breed of fiber lasers, BRFLs have shown unique spectral dynamics and noise properties, arousing immense potentials in underlying fundamental research as well as practical applications[48]. Providing that the linewidth of the BRFL is unknown, the acquisition of the delay-time resolved phase jitter variance was first deployed with variable trace time intervals from 0.1 ms to 10 ms. According to Eq. (3), the linewidth can be mathematically estimated as the slope efficiency by a linear curve fitting of the phase jitter versus the modified delay time, as shown in Fig.7 (a). By plotting the estimated linewidth as a function of the trace time interval, two areas can be divided in terms of the relation between the trace time interval $\Delta t$ and the laser coherent time ($t_c \cong 1/\Delta v_c$), as depicted in Fig. 7(b). Taking account of the requirement of the trace time interval ($\Delta t \geq t_c$) in the statistical retrieval of the phase jitter, the ultimate linewidth can be reasonably determined at the value close to the crossing point with a virtual curve of $t = 1/\Delta v_c$. As a result, the linewidth of the BRFL was given by the least square method-based linear fitting slope (i.e. $554.0 \pm 10.0$ Hz) of the delay-time resolved phase jitter variance at the trace interval of 1.8 ms. In Fig. 7(c), a normalized lorentzian PSD of the laser spectrum was re-established by substituting $\Delta v_c$ with the measured linewidth of 554 Hz in our proposal, compared to the measured results ($675\pm100$ Hz) via the DSHI with 200-km delay fibre which is estimated with the 20-dB linewidth (13.5 kHz) divided by 20 (See **Methods**). Evidently, the conventional DSHI with hundreds of kilometre fibre delay inevitably imposed a deviation of the intrinsic lorentzian linewidth with a Gaussian-like broadening in the central of the laser shape, even though corrected by the 20-dB linewidth rather than a direct 3-dB linewidth. Furthermore, the dependence of the pump laser on the BRFL linewidth was explicitly investigated by exploiting the delay-time-resolved phase jitter variance. As shown in Fig. 7(d), a 3.4-kHz narrow-linewidth pump laser delivers the random laser emission with a lowest phase jitter variance as well as a narrowest sub-kHz intrinsic linewidth while the pump lasers with larger linewidth would also remarkably deteriorate the phase noise of the random laser radiation and readily broaden its intrinsic linewidth beyond 1 kHz. It indicates the feasibility and the robustness of our proposal as an effective approach for exploiting underlying laser physics.

## Discussions

We demonstrate a technique relying on COTDR system to retrieve the delay-time-resolved phase noise dynamics of a coherent laser, yielding a high-precision determination of its intrinsic linewidth subject to the quantum limit. The delay-time-resolved phase jitter variance was deployed with up to dozens of microsecond delay time in the aid of Rayleigh scattering along few-kilometre-long fibres, which discriminately mitigates the dominated $1/f$ frequency noise over a long delay time (e.g., >1ms). Revealing of $\pi$-beyond phase jitter over a longer delay time, which currently is restricted by the PCC-based retrieved phase confinement within [0 $\pi$], however, can be computationally solved by the phase unwrapping algorithm[49]. In addition, alternative possibilities in terms of fading-diminishing phase retrieval approaches, including I/Q demodulation[50,51] and $3\times3$ coupler-based interferometer[52] merged with polarization diversity detection[53], would definitely diversify the options for a high-fidelity characterization of the intrinsic linewidth.

Considering the fact that slow phase fluctuations can be technically compensated[8,54], noise spectra at low frequencies are not always of concern while the intrinsic linewidth characteristic associated with fundamental limits is non-trivially paid extensive attentions; for instance, the intrinsic linewidth governed by diverse phase noise dynamics were intensively studied in a plenty of lasers involving various gain mechanisms such as gas laser[55], fibre laser[56], semiconductor laser[10,39], quantum cascade laser[9,32], photonic integrated laser[13] and micro-resonator laser[57]. To demonstrate the different phase noise dynamics of lasers with various gain mechanism, we have experimentally measured the delay-time-resolved phase jitter variance of a fiber laser compared to that of an external cavity semiconductor laser, which are presented in Supplementary Figure S1. It is believed that the proposed intrinsic linewidth determination via a statistical analysis of laser phase noise dynamics offers salient robustness in terms of reliability and flexibility, opening new windows for discovering the next frontiers of laser physics and measurement science.

## METHODS

**Coherent optical time domain reflectometry.** Coherent optical time domain reflectometry (COTDR) is a ubiquitous technique that typically captures the beating notes of a *cw* light source and its pulse-modulated signal from Rayleigh scattering along an optical fibre for a time-of-flight measurement of position-resolved response from external disturbance. Either intensity[58] or phase[43,50] interrogation based on COTDR could be implemented to achieve a distributed acoustic sensing.

**Delayed self-heterodyne interferometer approach.** In comparison, a conventional delayed self-heterodyne interferometer (DSHI) was utilized to evaluate the laser linewidth. The laser beam was launched into a fibre-based Mach-Zehnder interferometer which was deployed by utilizing an acousto-optic modulator with a 40-MHz carrier frequency shift in one arm and a long delay fibre of SMF in the another arm. Sufficiently long delays (i.e., larger than laser coherent time) are basically required to decorrelate two beams, and the detected power spectral density of the beating signal turns to a self-convolution of the laser spectrum and then the laser linewidth can be determined. In our experiments, the heterodyne beating signal was converted by an AC-coupled photodetector (PDB450CAC, Thorlabs) and its power spectral density (PSD) is displayed by a signal analyzer (FSW50, R&S). To eliminate the 1/f frequency noise induced Gaussian linewidth broadening, the lorentzian linewidth of a coherent laser based on DSHI method is practically estimated by the width at 20-dB lower than the maximum of the measured PSD instead of the conventional 3-dB full-width-half-maximum linewidth.

**Acknowledgements**

We thank Chengxian Zhang for loaning the 2-kHz External Cavity Laser and the assistance during the experiments. We also acknowledge the financial support of the Canada Research Chairs (CRC) program (no. 950231352); the Natural Sciences and Engineering Research Council of Canada (RGPIN-2015-06071).

**Author contributions**

L.Z. conceived and performed the experiments as well as the data analysis. L.Z. and L.C. conducted analytical calculations and numerical simulations. L.Z. prepared the figures and wrote the manuscript. X. B. supervised the project. All authors contributed to the fruitful discussion and the revision of the manuscript.

**Data availability.** The data that support this work are available on request from the corresponding author.

**Competing interests.** The authors declare no competing interests.

See **Supplementary information** for supporting content.